\documentstyle[12pt]{article}
  \def\lsim{\raise0.3ex\hbox{$<$\kern-0.75em\raise-1.1ex\hbox{$\sim$}}}
\def\gsim{\raise0.3ex\hbox{$>$\kern-0.75em\raise-1.1ex\hbox{$\sim$}}}
\def\noi{\noindent}

\def\bea{\begin{eqnarray}}  \def\eea{\end{eqnarray}}
\def\beq{\begin{equation}}   \def\eeq{\end{equation}}

\def\beeq{\begin{eqnarray}} \def\eeeq{\end{eqnarray}}

\begin{document}
\begin{center}
\vbox to 1 truecm {}
{\large \bf Multiparticle production in small-x deep inelastic} \\ 
{\large \bf  scattering and the nature of the Pomeron} \\

\vskip 8 truemm
{\bf A. Capella, A. Kaidalov, V. Neichitailo and J. Tran Thanh Van}\\

Laboratoire de Physique Th\'eorique et Hautes Energies\footnote{Laboratoire associ\'e
au Centre National de la Recherche Scientifique - URA D0063} \\
Universit\'e de Paris XI, B\^atiment 210,
F-91405 Orsay Cedex, France \\
\end{center}

\vskip 1 truecm

\begin{abstract}
Properties of multiparticle final states produced in the central rapidity region of small-$x$ deep
inelastic scattering (DIS) are discussed. It is pointed out that these properties contain
important information on the nature of the Pomeron - used for the description of
$\sigma_{\gamma^*P}^{(tot)}$ and diffractive production in DIS. It is shown that
models based on universality of the Pomeron predict charged particle multiplicities in small-$x$
DIS which are in good agreement with recent HERA results.  \end{abstract}

\vskip 2 truecm

\noi LPTHE Orsay 97-58\par
\noi November 1997 \par
\newpage
\pagestyle{plain}
\section{Introduction} 
\hspace*{\parindent}
Small-$x$ DIS provides a good testing ground for theoretical approaches to high-energy interactions
of photons and hadrons. A strong increase of the structure function of the proton $F_2(x, Q^2)$ as
$x$ decreases observed at HERA \cite{1r,2r} has been often considered as an indication of the
existence of the ``hard'' Pomeron, predicted by QCD-perturbation theory \cite{3r,4r}. On the other
hand models based on the standard perturbative QCD-evolution provide a good description of
experimental data \cite{5r}-\cite{8r}. However, these two approaches do not explain the striking
difference between the energy dependence of total cross sections for hadronic (or $\gamma p)$
interactions and $\gamma^*p$-interactions observed experimentally. In ref. \cite{9r} a
unified approach to $hp$, $\gamma p$ interactions and DIS has been proposed, where the Pomeron pole
is assumed to be universal, i.e. the same in soft and hard processes, with an intercept $\alpha_P(0)
\approx 1.2$. The non-universality of the energy behaviour for $hp(\gamma p)$ and
$\gamma^*p$-interactions is attributed to the different sizes of unitarity corrections 
(multipomeron exchanges) in these two classes of processes. It was argued that in DIS the
unitarization corrections due to multipomeron exchange are much smaller than in hadronic
interactions. In the latter the unitarity corrections reduce the effective Pomeron intercept from
$\alpha_P(0) \approx 1.2$ to roughly 1.1. On the contrary, in DIS the ``bare'' intercept
$\alpha_P(0) = 1.2$ is only slightly modified by the unitary corrections and provides a ``singular''
$\left ( F_2(x, Q^2) \sim x^{-0.2} \right )$ initial condition for perturbative QCD
evolution. The change between an effective intercept of about 1.1 and the bare one of 1.2 occurs
very quickly when $Q^2$ increases from zero (at $Q^2 \ \gsim \ 1$ GeV$^{\rm 2}$ the effective
intercept is already quite close to the bare one). \par

A study of structure functions of the proton can hardly discriminate between different
theoretical approaches. Investigation of diffractive production of hadrons by real and virtual
photons provides extra tests of different theoretical models. \par

Here we would like to point out that the properties of multiparticle production in DIS are
related to the nature of the Pomeron and can be used to test theoretical models. The same
situation takes place in high-energy hadronic interactions. A study of only
$\sigma_{hp}^{(tot)}(s)$ and $d\sigma^{(el)}_{hp}(s,t)/dt$ does not allow to
discriminate between models with large unitarization effects \cite{12r,13r} and models where these
effects are small \cite{14r}. However a fast increase with energy of charged particle densities in
the central rapidity region, broad multiplicity distributions and long-range rapidity correlations
observed in hadronic interactions clearly demonstrate the importance of effects related to the
multipomeron exchanges \cite{12r} \cite{15r}. \par

We shall demonstrate below that the model based on universality of the Pomeron, discussed above,
gives parameter free predictions for charged particle densities and multiplicities in the central
rapidity region of DIS. These predictions are in a good agreement with HERA experimental data
\cite{16r}.

\section{Description of the model} 

\hspace*{\parindent} The CKMT-model \cite{9r} for small-$x$ DIS corresponds to the diagrams for
elastic $\gamma^*p$-amplitude, shown in Fig. 1, where the upper blobs 
denote all diagrams which describe the coupling of the Pomeron to virtual photons. They contain in
particular diagrams corresponding to QCD-evolution. Asymptotically
(i.e. for $Y \equiv \ell n {W^2 \over s_0} \to \infty )$, at fixed $Q^2$, the Pomeron amplitude
increases as $\exp (\Delta y)$ (where $\Delta = \alpha_P(0) - 1)$ with increasing energy,  while the
blobs in Fig. 1 have finite widths in rapidity. The $s$-channel cutting of the diagrams of Fig. 1
corresponds to the natural physical picture of $\gamma^*P$ inelastic interaction \cite{17r}, shown
schematically in Fig. 2. It illustrates the fact that a highly virtual photon produces a cascade of
partons with virtualities decreasing along the chain. Systems of partons with small virtuality (of
large size) interact strongly with the proton either inelastically (Fig. 2a) or diffractively (Fig.
2b). This picture is similar to the models of Bjorken \cite{18r} and B\"uchmuller \cite{19r}. The
inelastic production amplitude in Fig. 2a can be expressed as a sum of inelastic cuttings of a
single Pomeron diagram (Fig. 1a), two Pomeron exchange diagram (Fig. 1b), etc. Some of these cuttings
are shown in Figures 3.\par

In the models based on the $1/N$-expansion, the Pomeron corresponds to diagrams for elastic
scattering with the topology of a cylinder. The cutting of Fig. 3a leads to a 2-chains
configuration as shown in Fig. 4a and that of Fig. 3c to the 4-chains configuration of Fig. 4b.
\par

The weights with which different configurations contribute to the multiparticle production are
given by AGK-cutting rules \cite{20r} and thus can be fixed if the relative contributions of
rescatterings in Fig. 1 are known. \par

It was already mentioned above that contributions of rescatterings in DIS are much smaller than in
hadronic interactions and in first approximation, at a\-vai\-la\-ble energies, it is possible to
consider only the $PP$-contribution. In this case, the latter contribution can be determined from
HERA data on large rapidity-gap events \cite{21r,22r}. The ratio of the second rescattering
($PP$-exchange) to $P$-exchange is equal (with minus sign) to $F_{2}^D/F_{2}$ (where $F_2^D$ is the
diffractive contribution to $F_2$), which is equal to 10 $\div$ 15 $\%$ in the region $10^{-4} < x <
10^{-2}$ and 2 GeV$^{\rm 2} \ \lsim \ Q^2 \ \lsim \ 10^2$ GeV. For hadronic and $\gamma p$
interactions, this ratio is substantially larger.\par

It is known that particular multiparticle configurations are more sensitive to contributions of
rescatterings than the total cross section. For example the shadowing contribution to the cutting
of a single Pomeron shown in Fig. 3b is four times larger than the contribution of the diagram
of Fig. 1b \cite{20r}. For a study of multiparticle production in DIS we will assume a definite
model for the series of rescatterings shown in Fig. 1. An adequate description of hadronic
interactions was obtained in the ``quasi-eikonal'' model \cite{23r}. It is the simplest
generalization of the eikonal model which allows one to calculate all rescatterings in terms of the
double rescattering term. So we shall apply the same model to DIS. \par

In this model the total cross section $\sigma_{\gamma^*p}(W^2, Q^2)$ can be written in the form
\cite{24r}

\beq
\label{1e}
\sigma_{\gamma^*p}^{(tot)}(W^2, Q^2) = \sum_{k=0}^{\infty} \sigma_{\gamma p}^k (W^2, Q^2) =
\sigma_P(W^2, Q^2) f\left ( {Z \over 2} \right ) \eeq

\noi where $k$ is the number of cut Pomerons ($k = 0$ corresponds to diffractive production). The
contribution of the single Pomeron exchange to $\sigma_{\gamma^*p}^{(tot)}$ is denoted by
$\sigma_P(W^2, Q^2) = g(Q^2) \exp (\Delta \xi )$ with $\Delta \equiv \alpha_P(0) - 1$ and $\xi
\equiv \ell n {W^2 \over Q^2} = \ell n {1 \over x}$. Here we have neglected the real part of the
Pomeron amplitude - which has a small effect on multiparticle production. The function $f(Z)$,
defined as

\beq
\label{2e}
f(Z) = \sum_{n=1}^{\infty} {(-Z)^{n-1} \over n \cdot n~!} \quad , 
\eeq 

\noi takes into account all rescatterings in terms of only one function  $Z = {C(Q^2) \over R^2 +
\alpha '_P \xi} \exp (\Delta \xi )$. The function $Z$ can be determined from the data on the
diffractive cross section

\beq
\label{3e}
\sigma_{\gamma^*p}^0 = \sigma_P (W^2, Q^2) \left [ f\left ( {Z \over 2} \right ) - f(Z) \right ]
\eeq

\noi or, more precisely, from the ratio

\beq
\label{4e}
R^D \equiv {\sigma_{\gamma^*p}^0 \over \sigma_{\gamma^*p}^{(tot)}} = {F_2^D(x, Q^2) \over F_2(x,
Q^2)} = \left [ 1 - {f(Z) \over f\left ( {Z \over 2} \right ) } \right ] \quad . \eeq

\noi The quantity $(R^2 +
\alpha '_P \xi )$ in $Z$ is related to the $t$-dependence of the diffractive production amplitude
with $R^2 = 2.2$ GeV$^{-2}$ and $\alpha '_P = 0.25$ GeV$^{-2}$ \cite{9r}. \par

The cross sections with $k$-cut Pomerons (corresponding to final configurations with $2k$-chains)
have the following form \cite{24r}

\beq
\sigma^k_{\gamma^*p} (W^2, Q^2) = {\sigma_P \over kZ} \left [ 1 - \exp (- Z) \sum_{i=0}^{k-1} {Z^i
\over i !} \right ] \quad , \quad i \geq 1 \ \ \ . 
\label{5e}
\eeq

It follows from eqs. (\ref{2e})-(\ref{4e}) that for small values of the ratio $R^D$ one has $Z
\approx 8 \cdot R^D$. Experiments on diffraction dissociation at HERA show that the ratio $R^D$ is
practically $Q^2$-independent and we shall consider $C(Q^2)$ as a constant in future calculation.
From the HERA data on the ratio $R^D$ \cite{21r,22r,9r} we obtain $C \simeq 1.5$ GeV$^{\rm -2}$. \par

In order to calculate multiparticle production in the chains of Fig. 4 we shall use the DPM-model
\cite{12r} or QGSM-model \cite{13r} which give a good description of rapidity and multiplicity
distributions in hadronic interactions. Thus we do not introduce any new free parameter in the
calculation of multiparticle production in DIS. \par

In both models the hadronic spectra of each chain are obtained from a convolution of momentum
distribution and fragmentation functions. The former are obtained from Regge intercepts \cite{12r}
\cite{13r}. In the QGSM, used in the following calculations, the fragmentation functions into
charged hadrons are given in ref. \cite{25r} with a normalization constant $a_h = 1.14$. The
multiplicity distributions are computed assuming a Poisson distribution in clusters for the
individual chains - with fixed rapidity positions of the chain ends. The formalism is described in
detail in ref. \cite{12r}. The only parameter is the average charged multiplicity of the cluster
decay, for which we use the same value $K = 1.4$ as in previous works. The weights for the
configurations, with $k$-cut Pomerons ($2k$-chains) are given by eq. (\ref{5e}). (A Monte Carlo code
for multiparticle production in $\gamma p$ collisions, based on DPM, has been introduced by Engel
and Ranft \cite{26r}). \par

It is important to note that our calculations are reliable only in the central rapidity region and
the proton fragmentation region, while the fragmentation region of the virtual photon can be
dominated by perturbative production of jets of the type shown in Fig. 2. This region occupies an
average rapidity $\sim \ell n {Q^2 \over m^2_{\bot h}}$, out of the total rapidity
interval $\sim \ell n {W^2 \over m^2_{\bot h}}$. Unfortunately experiments at HERA cover mostly the
photon fragmentation region and only a part of a central rapidity region. In the following
comparison with experiment we will consider only the data in the central region of rapidity. 

\section{Comparison with experiment} 
\hspace*{\parindent} Let us consider first the density of charged hadrons in the central rapidity
region ${dn^h (W^2, Q^2, y) \over dy} = {1 \over \sigma^{(tot)} (W^2, Q^2)} {d \sigma^h(W^2, Q^2,
y) \over dy}$. It is convenient also to consider the corresponding quantity for nondiffractive
processes ${dn^h_{ND} \over dy} = {1 \over \sigma^{ND}} {d \sigma_{ND}^h \over dy}$, where
$\sigma^{ND} \equiv \sigma^{(tot)} - \sigma^0 = \sum\limits_{k=1}^{\infty} \sigma^k$. Due to
AGK-cancellations \cite{20r} the inclusive cross section in the central rapidity region
${d\sigma^h \over dy} \approx {d \sigma_{ND}^h \over dy}$ is determined only by the contribution
of the Pomeron pole and thus behaves as $\left ( {1 \over x} \right )^{\Delta}$. For hadronic
interactions, where rescattering effects for $\sigma^{(tot)}$ or $\sigma^{ND}$ are very important,
the increase of $\sigma^{(tot)}$ with energy is much slower and, as a result, ${dn^h \over dy}$
(as well as ${dn_{ND}^h \over dy}$) have a fast increase with energy. It was already emphasized in
ref. \cite{27r} that in DIS $\sigma^{(tot)}$ and $\sigma^{ND}$ have approximately the same $\left
( {1 \over x} \right )^{\Delta}$-behaviour as ${d\sigma \over dy}$ (due to the small effect of the
rescatterings) and ${dn^h \over dy}$ should have weaker energy dependence than for hadronic
interactions. Likewise, multiplicity distributions in DIS are expected to be narrower
than in $pp$ or $\gamma p$ and to show a much smaller KNO-scaling violation.\par

Predictions of the model for the mean charged multiplicity of nondiffractive events in the
central pseudorapidity domains of DIS as functions of energy ($W$) are compared with H1-data
\cite{16r} in Fig. 5a. Densities of particles in chains are taken from the model \cite{13r}. The
DPM model leads to practically identical results. As explained above, the predictions of the model
are most reliable for the smallest pseudorapidities ($1 < \eta^* < 2$), where they are in good
agreement with experiment both in absolute magnitude and in the rate of increase with energy. In
Fig. 5a we also show the results for the region $1 < \eta^* < 3$ where the agreement between theory
and experiment, although still reasonable, is less good. It becomes slightly worse when the size of
the rapidity interval increases. Note that the agreement in the range $1 < \eta^* < 3$ is better at
higher energies, where the distance between central rapidity region and fragmentation region of the
virtual photon is larger. Model predictions for the rapidity distributions in DIS (for $Q^2 =$ 15
GeV$^{\rm 2}$) at different energies are shown in Fig. 5b. A substantial increase of the central
plateau is predicted. At energies $W \lsim$ 200 GeV this increase is mainly connected with the
increase of charged particle densities in the Pomeron chains. The faster increase at energies
$\sqrt{W} > 10^3$ GeV is due to an increase of the average number of chains (cutted Pomerons). At
these extremely high energies ($x \ \lsim \ 10^{-5}$) effects of unitarization for structure
functions are very important. Note, that as anticipated above, the energy rise of the central
plateau in DIS is substantially smaller than the one measured in $\bar{p}p$ collisions. \par

The experimental data for charged hadrons produced in nondiffractive
DIS in different pseudorapidity intervals of the central region \cite{16r} are compared with our
predictions in Figs. 6. The agreement with experiment is good in the interval $1 < \eta^* < 2$ and
becomes worse for larger intervals). \par

Comparison of theoretical predictions for different moments
of multiplicity distributions with experimental data \cite{16r} is given in Table~I. It is important
to notice that the values of these moments are substantially smaller than the ones measured in
$\bar{p}p$ scattering at the same energy (see Ref. \cite{28r}). This is a clear confirmation of the
main feature of our approach, namely that the size of the rescattering contributions is smaller in
DIS than in $pp$. \par

A further confirmation of this result could be obtained from the study of the energy dependence
of the moments of the multiplicity distribution in DIS. It has been predicted in \cite{29r} that in
the central rapidity interval $|y^*| < 1.5$, the moments $C_i$ in $\bar{p}p$ should have a strong
energy dependence. This dependence has been confirmed experimentally \cite{28r} (see also
\cite{12r}). Such an energy dependence is due to the increase with energy of the fluctuations in the
number of chains. Since the multi-chain (i.e. multi-cut Pomeron) contributions are smaller in DIS,
we expect a smaller increase of the moments with energy. Note that models such as JETSET and MEPS
based on perturbative QCD predict a magnitude and energy dependence of the DIS moments $C_i$ in the
central rapidity $1< \eta^* < 2$, which is substantially different from our predictions \cite{16r}.
\par

 Experimental studies of $p_{\bot}$-dependences of final hadrons at HERA \cite{30r,31r} are also
in agreement with our assumption that in the central rapidity region ($y^* < 3$, for $W = 200$ GeV or
$x_F < 0.07$) properties of final states are similar to the ones in $\gamma p$ or $hp$. Average
transverse momenta in this region are much smaller than in the fragmentation region of the virtual
photon \cite{30r}. \par

Finally, let us now comment on multiparticle production in the virtual photon fragmentation region,
where QCD perturbation theory can be applied. Though our model is strictly speaking not reliable in
this region it still reproduces some features of multiplicity distributions up to pseudorapidity
$\eta^* \sim 4$ though the agreement with experiment is worse than in the central
rapidity region $\eta^* < 2 \div 3$. The model predicts very weak dependence of ${dn \over d\eta^*}$
on $Q^2$ for fixed value of $W$ and a rather strong dependence on $W$ for fixed $Q^2$. These
consequences of the model are in an agreement with experimental results \cite{16r}, which show
that mean charged particle multiplicity is practically independent of $Q^2$ in the interval $10$
GeV$^{\rm 2} < Q^2 < 400$ GeV$^{\rm 2}$ for fixed $W$. This observation is not easy to reconcile
with predictions based on perturbative QCD, - radiation of extra gluons in the final state should
lead to a substantial increase of multiplicity as $Q^2$ increases \cite{32r}. 	It may be an
indication that the non-perturbative dynamics described above is important even in the
fragmentation region of the virtual photon. \par

\section{Conclusions} 
\hspace*{\parindent}
Our analysis indicates that production of hadrons in the central rapidity region in DIS can be
understood in an approach based on the universality of the Pomeron. The model presented here does
not contain any adjustable parameter. The size of the rescattering contribution has been
determined from diffractive production data in DIS and all other parameters are fixed \cite{12r},
\cite{13r} from high-energy hadronic interaction data. The model gives a quantitative description of
the densities of charged particles in the central rapidity region and their multiplicity
distributions at HERA energies. Possible tests of our approach in forthcoming experiments have
also been presented. \par

Our results provide a clear confirmation of the idea, introduced in previous works
\cite{9r}-\cite{10r}, that unitarity corrections, which play a capital role in $pp$ interactions, are
substantially smaller in DIS. Our model can provide a quantitative way of relating multiparticle
production in hard and soft processes.

\section*{Acknowledgements} 
\hspace*{\parindent}
It is a pleasure to thank Yu. Dokshitzer, V. Khoze, G. Korchemsky, A. Krzywicki, G. Marchesini, A.
Mueller and R. Peschanski for useful discussion. \par

This work was supported in part by grant INTAS 93-79 ext. and grant 96-02-19184 of the Russian
Foundation for Fundamental Investigations.

\newpage
\section*{Figure Captions}

\noindent {\bf \underbar{Fig. 1} :}\par
The Pomeron-exchange diagrams for elastic $\gamma^*p$-scattering amplitude. \par \vskip
1 truemm

\noindent {\bf \underbar{Fig. 2} :}\par
Inelastic cuttings of the diagrams of Fig. 1.  a) Totally inelastic final state.\break \noindent 
b) Diffractive dissociation of a virtual photon. \par \vskip 1 truemm

\noindent {\bf \underbar{Fig. 3} :} \par a) $s$-channel cutting of the single Pomeron exchange
diagram of Fig. 1a). \par
b), c) Totally inelastic cuttings of the $PP$-exchange diagram of Fig. 1b).
 \par \vskip
1 truemm

\noindent {\bf \underbar{Fig. 4} :}\par
a) Two chains configuration corresponding to the diagram of Fig. 3a). \par
b) Production of 4-chains
corresponding to the diagram of Fig. 3c).  \par \vskip 1 truemm

\noindent {\bf \underbar{Fig. 5} :}\par
a) Comparison of theoretical predictions for energy dependence of mean charged multiplicity of
nondiffractive events in central pseudorapidity domains of DIS with experimental data
\protect{\cite{16r}}. Lower points and curve correspond to $1 < \eta^* < 2$ and the upper ones to
$1 < \eta^* < 3$. \par

b) Model predictions for the rapidity distributions of charged particles in the central rapidity
region at various energies. \par \vskip 1 truemm

\noindent {\bf \underbar{Fig. 6} :}\par
Multiplicity distributions of charged hadrons produced in nondiffractive events in central
pseudorapidity domains of DIS \protect{\cite{16r}} compared to theoretical predictions. a) $W = 97$
GeV, $1 < \eta^* < 2$~; b) $W = 202$ GeV, $1 < \eta^* < 2$~; c) $W = 97$ GeV, $1 < \eta^* < 3$~; d)
$W = 202$ GeV, $1 < \eta^* < 3$. 

\section*{Table Caption}

\noindent {\bf \underbar{Table 1} :} \par

Model predictions for various moments and cumulants of the multiplicity distributions of charged
particles in DIS in two central pseudorapidity intervals and at two different energies are compared
with H1 data \cite{16r}.

\newpage
\begin{center}
\begin{tabular}{|c|cccc|}
\hline
&\multicolumn{4}{|c|} {$<W> = 201.9$}\\
&$1 < \eta^* < 2$ &$1 < \eta^* < 2$ &$1 < \eta^* < 3$ &$1 < \eta^* < 3$  \\ \hline
$C_2$ &1.69 &1.69 $\pm$ 0.02 $\pm$ 0.03 &1.48 &1.44 $\pm$ 0.01 $\pm$ 0.02 \\ 
$C_3$ &3.81 &3.75 $\pm$ 0.12 $\pm$ 0.20 &2.83 &2.62 $\pm$ 0.09 $\pm$ 0.09  \\ 
$C_4$ &10.73 &10.02 $\pm$ 0.65 $\pm$ 1.22 &6.68 &5.71 $\pm$ 0.47 $\pm$ 0.34  \\ 
$R_2$ &1.34 &1.32 $\pm$ 0.02 $\pm$ 0.03 &1.29 &1.26 $\pm$ 0.01 $\pm$ 0.03  \\ 
$R_3$ &2.26 &2.11 $\pm$ 0.09 $\pm$ 0.13 &2.06 &1.88 $\pm$ 0.08 $\pm$ 0.11 \\ 
$K_3$ &0.255 &0.147 $\pm$ 0.043 $\pm$ 0.101 &0.204 &0.108 $\pm$ 0.041 $\pm$ 0.036 \\ \hline
&\multicolumn{4}{|c|} {$<W> = 96.9$} \\
&$1 < \eta^* < 2$ &$1 < \eta^* < 2$ &$1 < \eta^* < 3$ &$1 < \eta^* < 3$ \\ 
\hline
$C_2$ &1.71 &1.71 $\pm$ 0.02 $\pm$ 0.02 &1.48 &1.40 $\pm$ 0.01 $\pm$ 0.03 \\
$C_3$ &3.85 &3.87 $\pm$ 0.16 $\pm$ 0.18 &2.78 &2.41 $\pm$ 0.04 $\pm$ 0.16 \\
$C_4$ &10.72 &10.67 $\pm$ 0.93 $\pm$ 1.23 &6.38 &4.83 $\pm$ 0.19 $\pm$ 0.64 \\
$R_2$ &1.31 &1.30 $\pm$ 0.02 $\pm$ 0.03 &1.26 &1.19 $\pm$ 0.01 $\pm$ 0.03 \\
$R_3$ &2.11 &2.11 $\pm$ 0.14 $\pm$ 0.15 &1.91 &1.64 $\pm$ 0.04 $\pm$ 0.13 \\ 
$K_3$ &0.139 &0.187 $\pm$ 0.070 $\pm$ 0.097 &0.139 &0.048 $\pm$ 0.015 $\pm$ 0.044 \\
\hline 
\end{tabular}
\vskip 3 truemm
{\bf Table 1}
\end{center}

\newpage
\begin{thebibliography}{99} \par \vskip 5 truemm 
\bibitem{1r} M. Derrick et al (ZEUS Collaboration), Phys; Lett. {\bf B293}, 465 (1992)~; {\bf B316},
41 (1993).  

\bibitem{2r} T. Ahmed et al (H1 Collaboration), Phys. Lett. {\bf B299}, 85 (1992)~; I. Abt et
al (H1), Nucl. Phys. {\bf B407}, 515 (1993).     

\bibitem{3r} V. S. Fadin, E. A. Kuraev and L. N. Lipatov, Phys. Lett. {\bf B60}, 50 (1975)~; Sov.
Phys. JETP {\bf 45}, 199 (1977)~; I. I. Balitsky and L. N. Lipatov, Sov. J. Nucl. Phys. {\bf 28},
822 (1978). 
  
\bibitem{4r} N. N. Nikolaev, B. G. Zahharov, J. Exp. Theor. Phys. {\bf 78}, 598 (1994)~; Phys.
Lett. {\bf B327}, 149, 157 (1994)~; N. N. Nikolaev, B. G. Zakharov and V. R. Zoller, J. Exp.
Theor. Phys. {\bf 78}, 806 (1994)~; Phys. Lett. {\bf B328}, 486 (1994)~; A. Mueller and H.
Patel, Nucl. Phys. {\bf B425}, 471 (1994).    

\bibitem{5r} M. Gluck, E. Reya and A. Vogt, Z. Phys. {\bf C53}, 127 (1992).
  
\bibitem{6r} A. D. Martin, R. G. Roberts and W. J. Stirling, Phys. Rev. {\bf D50}, 6734 (1994).
  
\bibitem{7r} H. L. Lai et al (CTEQ Collaboration), Phys. Rev. {\bf D51}, 4763 (1996).  
 
\bibitem{8r} R. D. Ball and S. Forte, Phys. Lett. {\bf B335}, 77 (1994)~; {\bf B336}, 77 (1994). 
  
\bibitem{9r} A. Capella, A. Kaidalov, C. Merino and J. Tran Thanh Van, Phys. Lett. {\bf B337},
358 (1994).

\bibitem{10r} A. Capella, A. Kaidalov, C. Merino and J. Tranh Thanh Van, Phys. Lett. {\bf B343},
403 (1995)~; A. Capella, A. Kaidalov, C. Merino, D. Pertermann and J. Tran Thanh Van, Phys. Rev.
{\bf D53}, 2309 (1996). 

 \bibitem{11r} V. N. Gribov, Zh. Eksp. Teor. Fiz. {\bf 57}, 654 (1967). 

\bibitem{12r} A. Capella, U. Sukhatme, C.-I. Tan and J. Tran Thanh Van, Phys. Rep. {\bf 236}, 225
(1994).

 \bibitem{13r} A. Kaidalov, in QCD at 200 TeV, eds. L. Cifarelli and Yu. Dokshitzer (Plenum, New
York, 1992) p. 1.  

\bibitem{14r} A. Donnachie and P. V. Landshoff, Nucl. Phys. {\bf B244}, 332 (1984)~; {\bf B267},
690 (1986).

  \bibitem{15r} A. Capella and A. Krzywicki, Phys. Rev. {\bf D18}, 4120 (1978)~; A. Kaidalov, Proc.
of XXII Intern. Conf. on Multiparticle Dynamics, ed. by C. Pajares, p. 185 (1992)~; 

\bibitem{16r} S. Aid et al (H1 Collaboration), DESY 96-160.  

\bibitem{17r} A. Kaidalov, Surveys in High Energy Physics {\bf 9}, 143 (1996).
 
\bibitem{18r} J. D. Bjorken and J. Kogut, Phys. Rev. {\bf D8}, 1341 (1973)~; J. D. Bjorken,
SLAC-PUB-7096. 

\bibitem{19r} W. B\"uchmuller, Phys. Lett. {\bf B353}, 335 (1995)~; W. Buchmuller and A. Hebecker,
Phys. Lett. {\bf 355}, 573 (1995)~; Nucl. Phys. {\bf B476}, 203 (1996).   


\bibitem{20r} V. Abramovskii, V. N. Gribov and O. V. Kancheli, Sov. J. Nucl. Phys. {\bf 18}, 308
(1974).

\bibitem{21r} M. Derrick et al (Zeus Collaboration), Phys. Lett. {\bf 315}, 481 (1993)~; {\bf
B332}, 228 (1994)~; {\bf B338}, 477 (1994)~; Z. Phys. {\bf C70}, 391 (1996).

 \bibitem{22r} T. Ahmed et al (H1 Collaboration), Nucl. Phys. {\bf B429}, 477 (1994)~; Phys. Lett.
{\bf B348}, 681 (1995).

 \bibitem{23r} K. A. Ter-Martirosyan, Nucl. Phys. {\bf B36}, 566 (1972).

\bibitem{24r} K. A. Ter-Martirosyan, Phys. Lett. {\bf 44B}, 377 (1973).

\bibitem{25r} A. Kaidalov, Yad. Fiz. {\bf 45}, 1452 (1987).

 \bibitem{26r} R. Engel, Z. Phys. {\bf C66}, 203 (1995)~; R. Engels and J. Ranft, ENSLAPP-A-540/95
(hep-ph/9509373).

\bibitem{27r} A. Kaidalov, Proc. of the Workshop on the Leading Particle Effect, Erice, October
(1996).

\bibitem{28r} UA5 collaboration, R. E. Ansorge et al., Z. Phys. {\bf C43}, 357 (1989).

\bibitem{29r} A. Capella, A. Staar and J. Tran Thanh Van, Phys. Rev. {\bf D32}, 2933 (1985)~; A.
Capella and J. Tran Thanh Van, Z. Phys. {\bf C23}, 165 (1984).

 \bibitem{30r} I. Abt et al (H1 Collaboration), Z. Phys. {\bf C63}, 377 (1994).

\bibitem{31r} M. Derrick et al (Zeus Collaboration), Z. Phys. {\bf C70}, 1 (1996).

\bibitem{32r} L. V. Gribov, Yu. L. Dokshitzer, V. A. Khoze and S. I. Troyan, JETP Lett. {\bf 45},
515 (1987)~; Sov. Phys. JETP {\bf 67}, 1303 (1988)~; Phys. Lett. {\bf B202}, 276 (1988).  
\end{thebibliography}
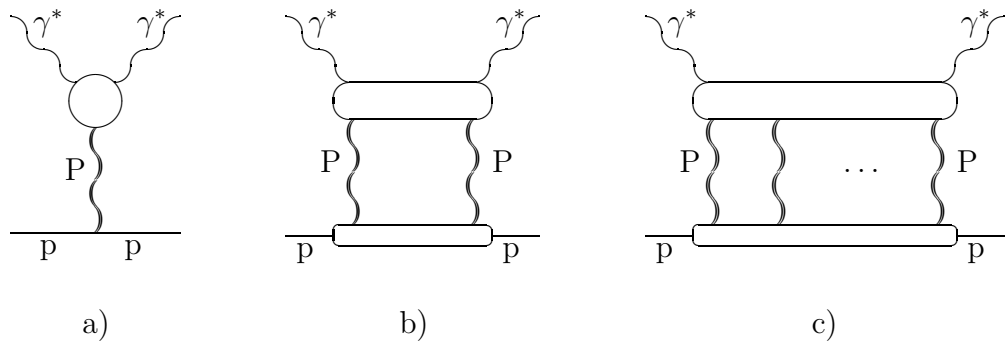
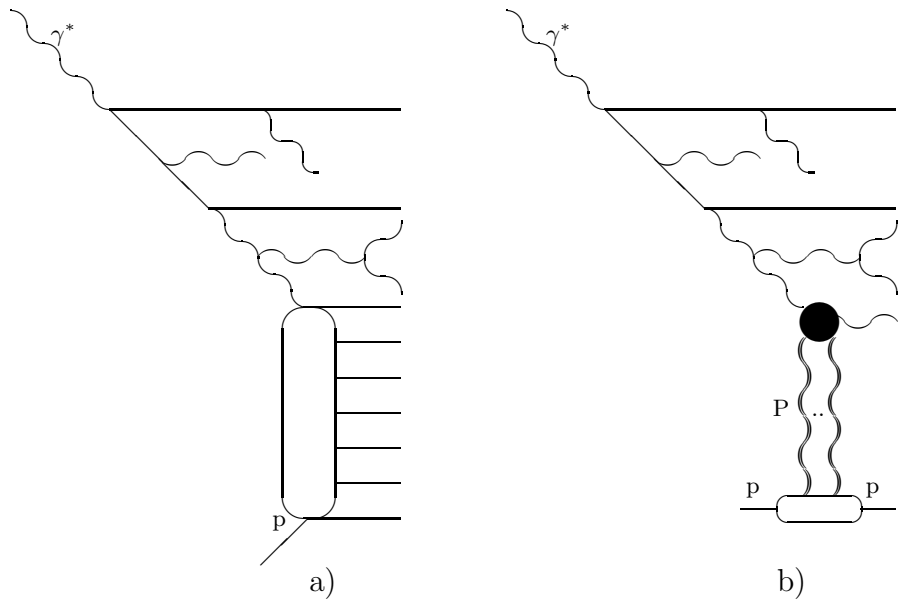
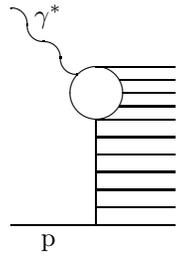
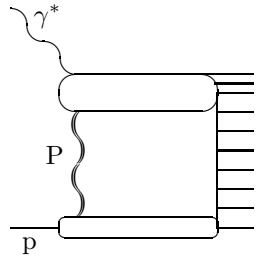
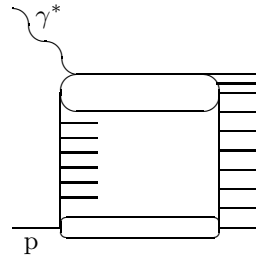
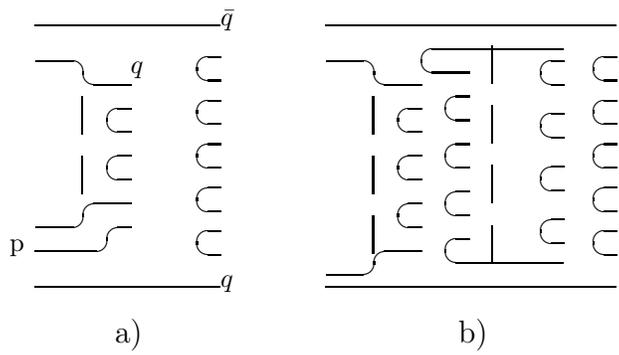
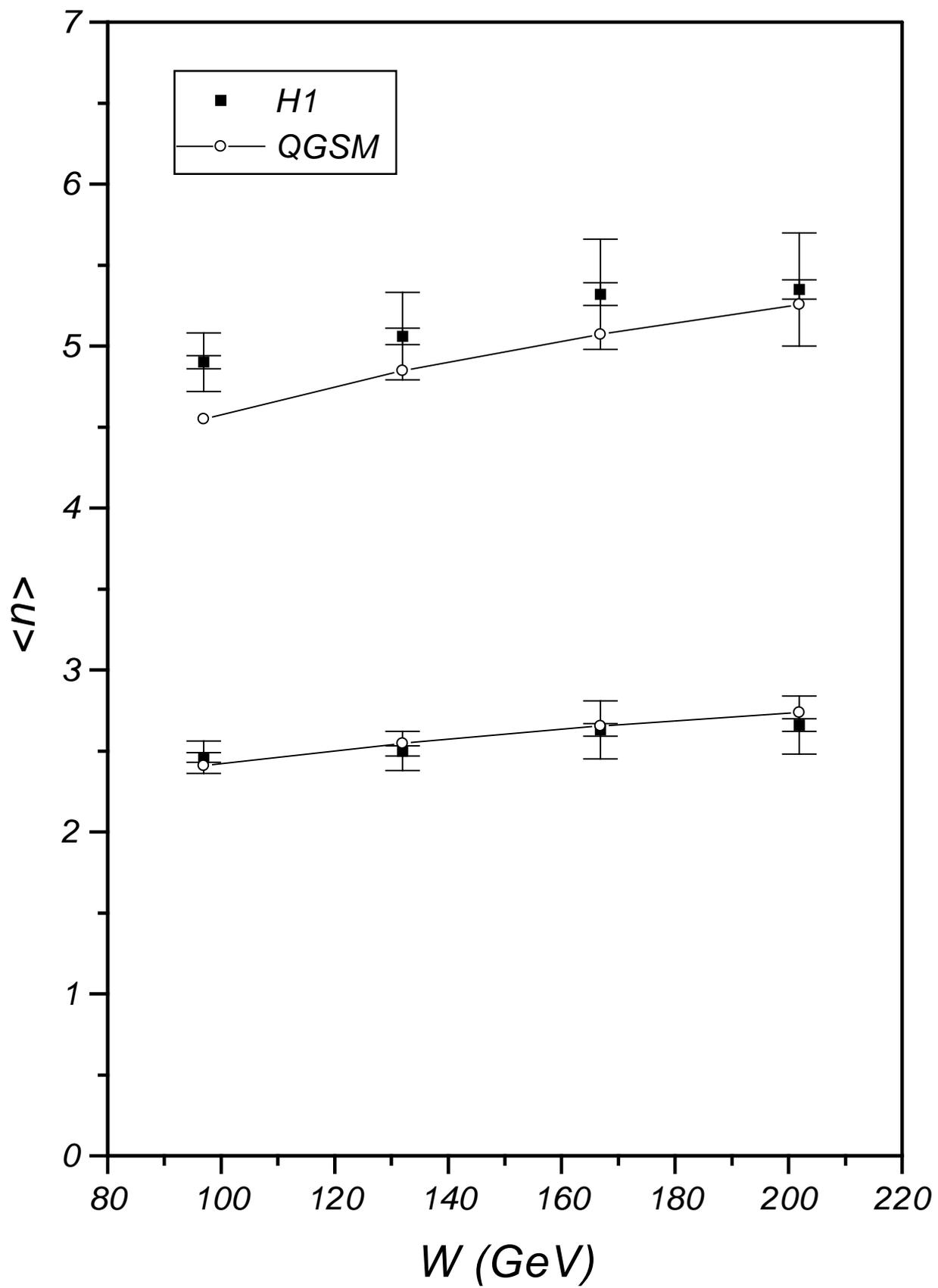
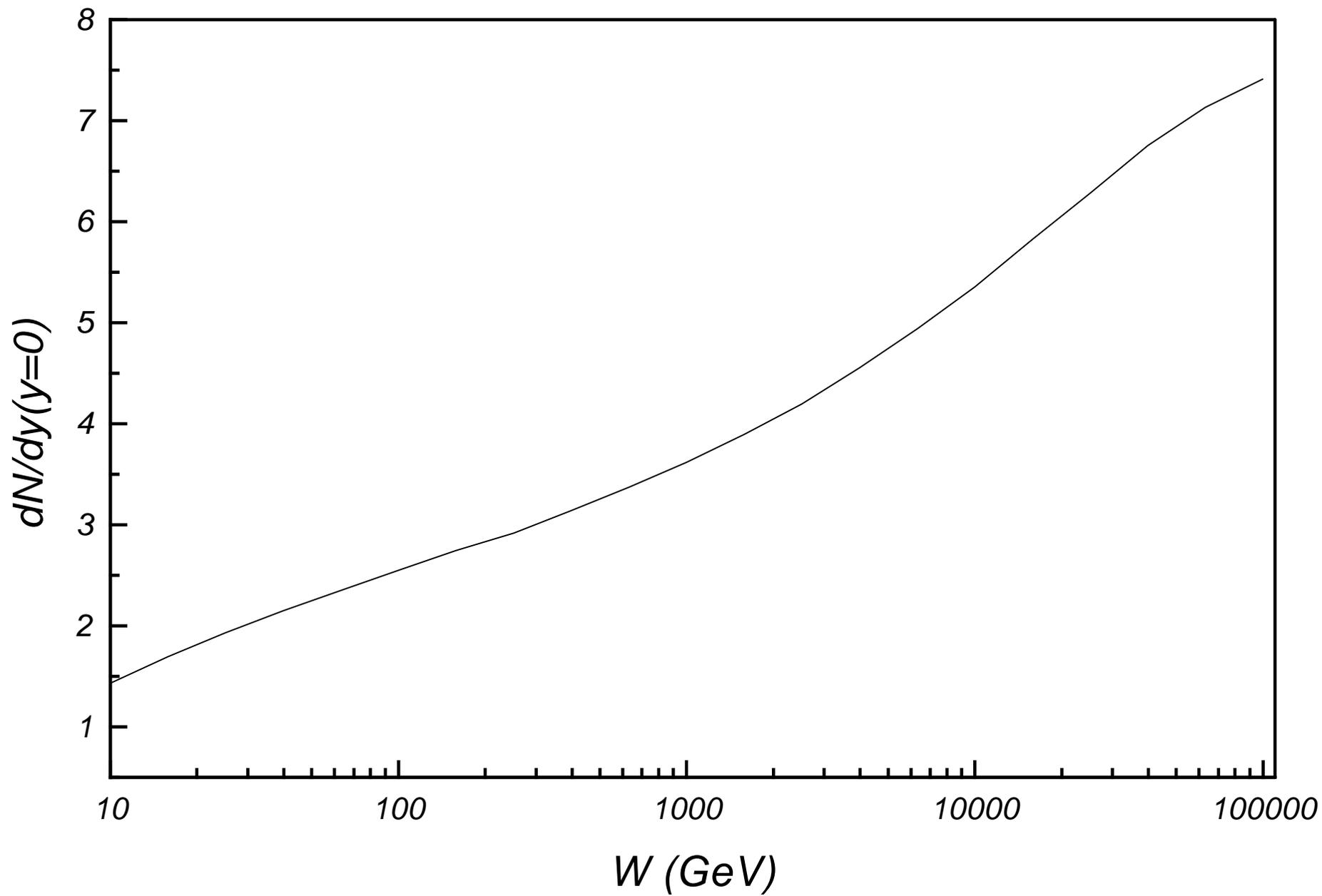
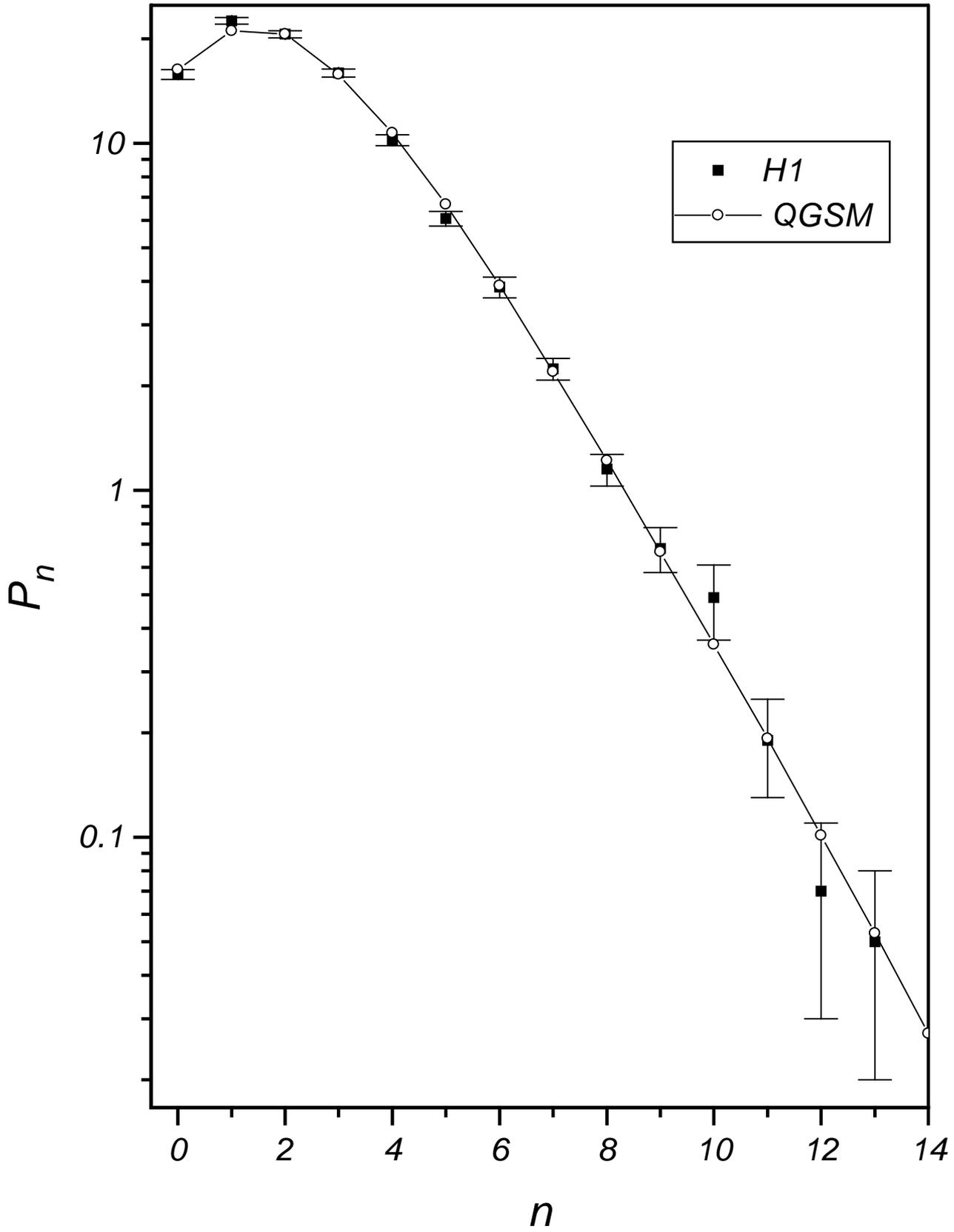
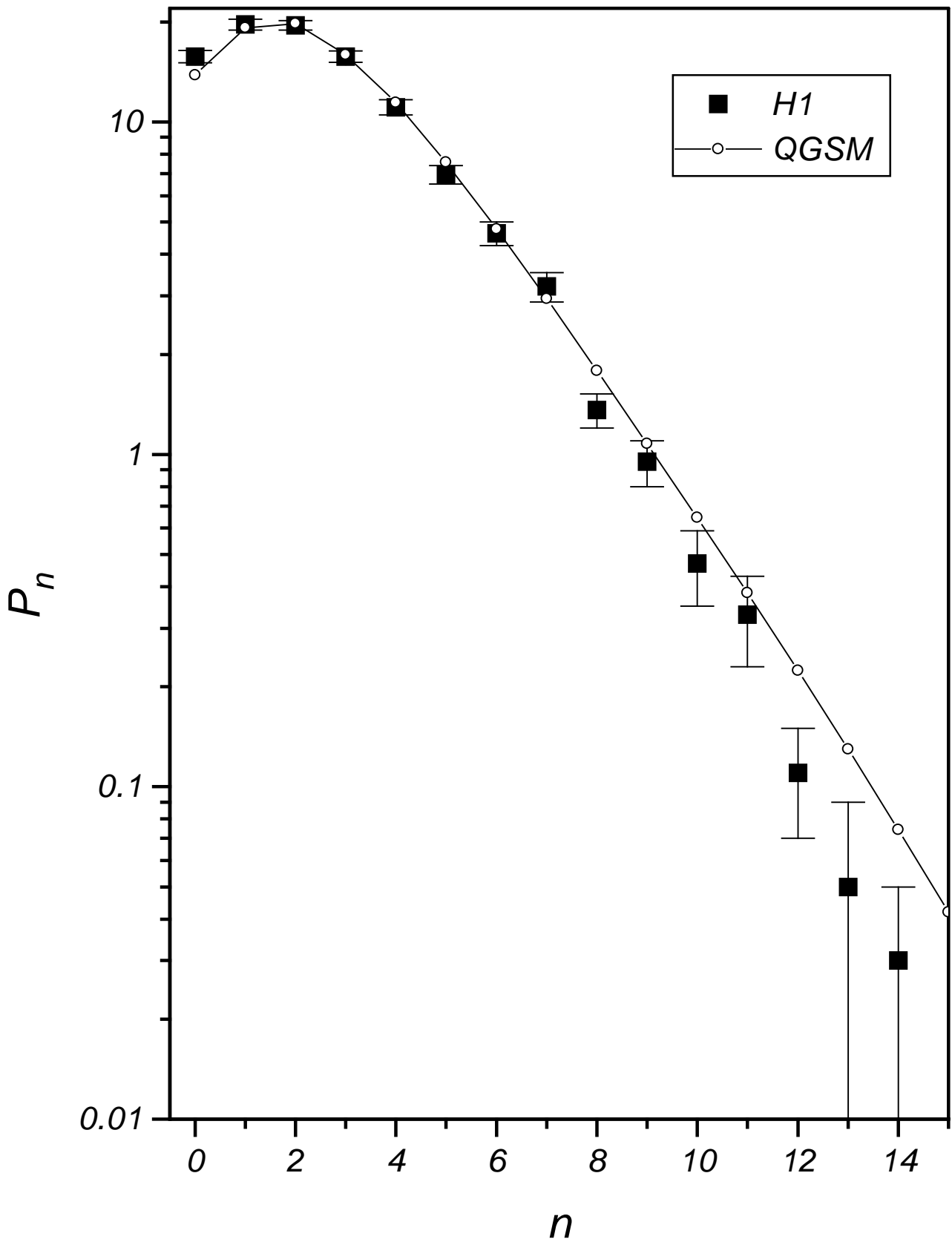
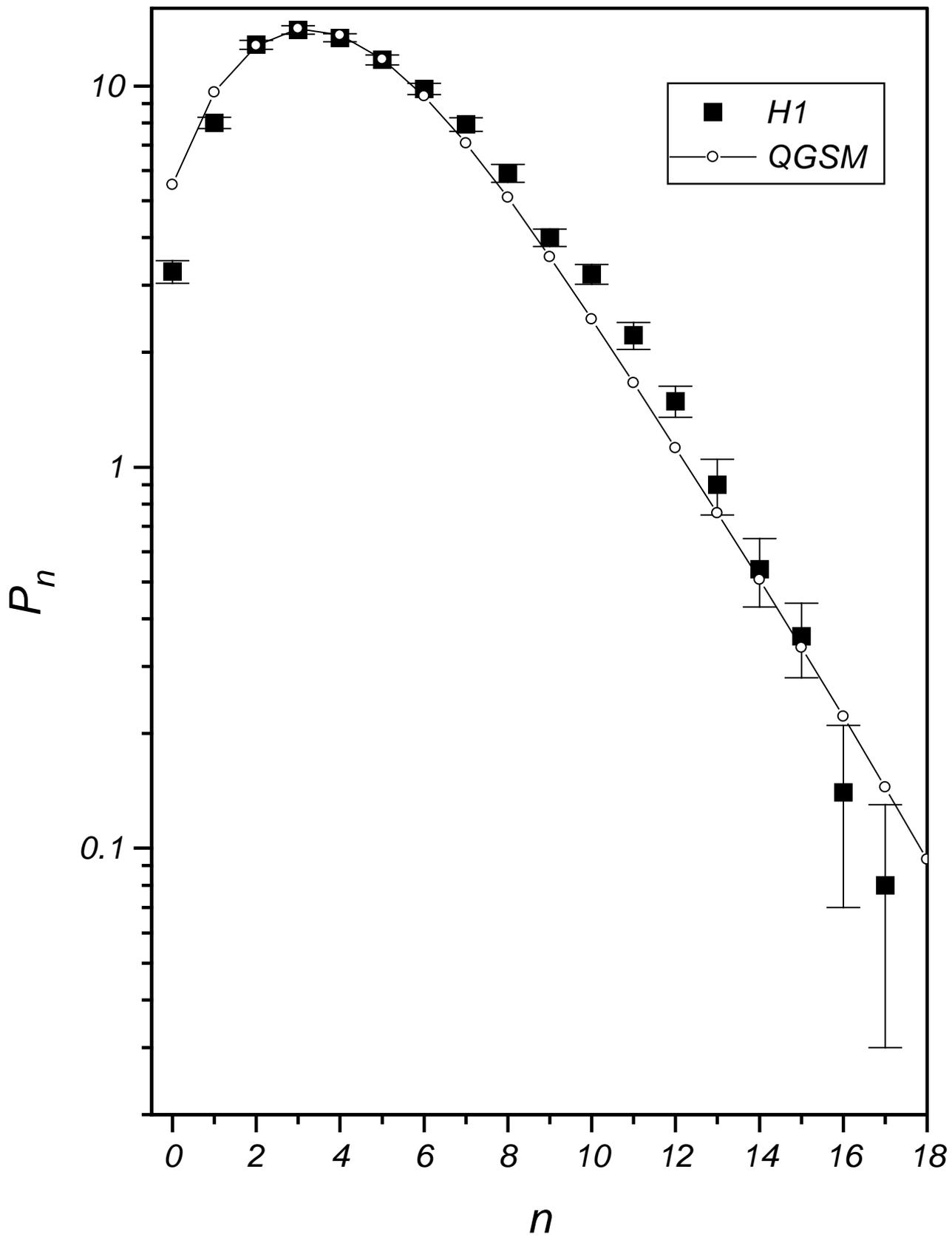
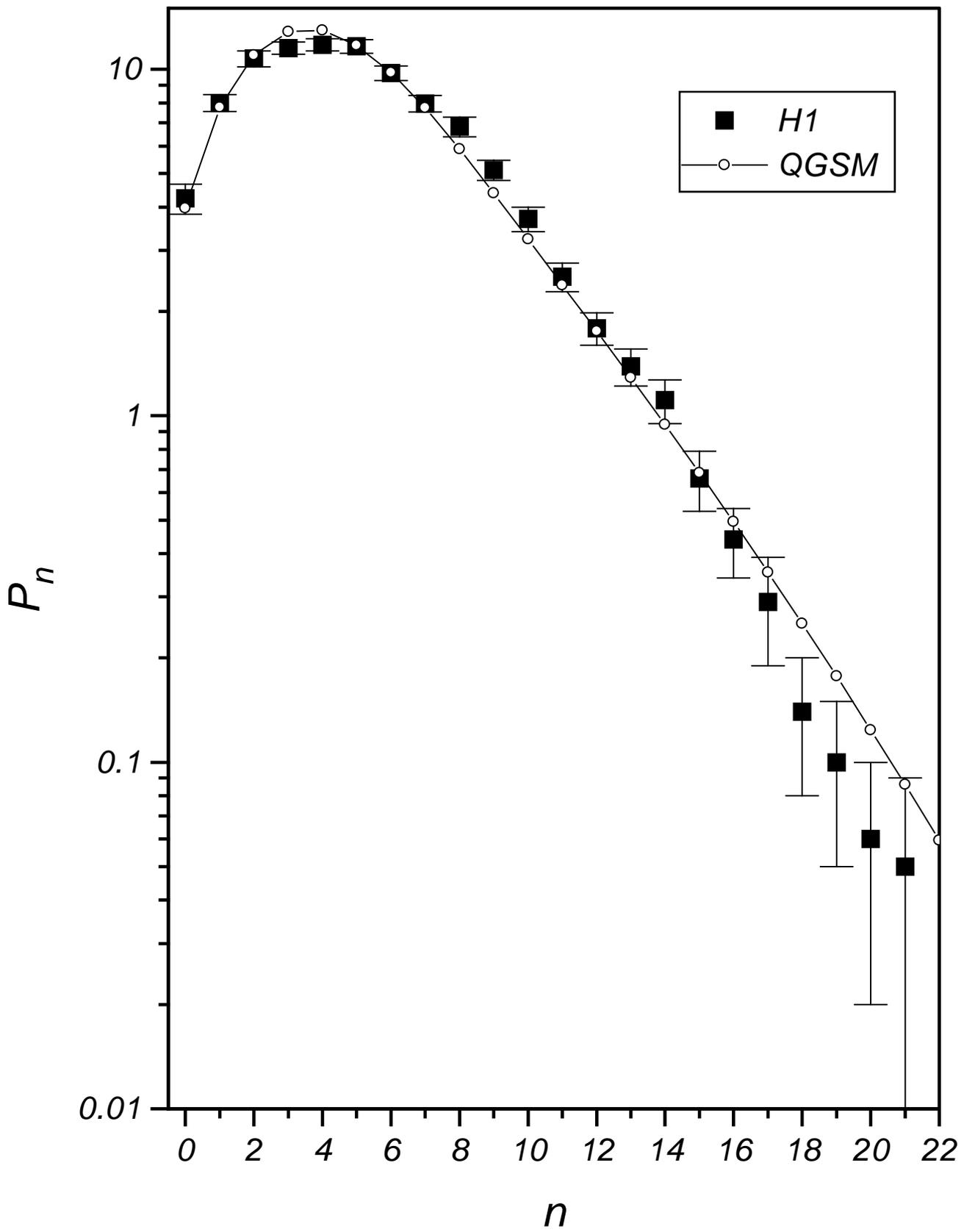
\end{document}